\documentclass[mathpazo]{cicpmik}
\usepackage{amsmath}
\usepackage{amssymb}
\usepackage{graphicx}
\usepackage{hyperref}
\usepackage[arrow, matrix, curve]{xy}
\usepackage{bm}
\usepackage{yhmath}
\usepackage{color}

\renewcommand{\vec}[1]{\mathbf{#1}}

\usepackage[usenames,dvipsnames]{xcolor}
\hypersetup{colorlinks=true, linkcolor=BrickRed, urlcolor=blue!50!black, citecolor=blue!50!black}

\makeatletter
\newcommand\hide@visible[1]{%
  \bgroup\fboxsep=.3ex\colorbox{Gray}{begin hide}%
  #1\colorbox{Gray}{end hide}\egroup%
}
\newcommand\hide@hidden[1]{%
  \bgroup\fboxsep=.3ex\colorbox{Gray}{hidden text}%
}
\newcommand\hide@invisible[1]{}
\newcommand\makevisible{\let\hide\hide@visible}
\newcommand\makehidden{\let\hide\hide@hidden}
\newcommand\makeinvisible{\let\hide\hide@invisible}
\makeatother
\makehidden

\newcommand{\lognhg}{{\it log-nonug}}
\newcommand{\nhg}{{\it nonug}}

\begin{document}
\title{An improved integration scheme for Mode-coupling-theory equations}

\author[Caraglio M et.~al.]{Michele Caraglio\affil{1}\comma\corrauth,
                            Lukas Schrack\affil{1},
                            Gerhard Jung\affil{1}, and
                            Thomas Franosch\affil{1}}
\address{\affilnum{1}\ Institut f\"ur Theoretische Physik, Universit\"at Innsbruck, Technikerstra{\ss}e 21A, A-6020, Innsbruck, Austria}
\emails{{\tt Michele.Caraglio@uibk.ac.at} (M.~Caraglio), 
        {\tt Lukas.Schrack@uibk.ac.at} (L.~Schrack), 
        {\tt Gerhard.Jung@uibk.ac.at} (G.~Jung), 
        {\tt Thomas.Franosch@uibk.ac.at} (T.~Franosch)}

%%%%% Begin Abstract %%%%%%%%%%%
\begin{abstract}
Within the mode-coupling theory (MCT) of the glass transition, we reconsider the numerical schemes to evaluate the MCT functional.
Here we propose nonuniform discretizations of the wave number, in contrast to the standard equidistant grid, in order to decrease the number of grid points without losing accuracy.
We discuss in detail how the integration scheme on the new grids has to be modified from standard Riemann integration.
We benchmark our approach by solving the MCT equations numerically for mono-disperse hard disks and hard spheres and by computing the critical packing fraction and the nonergodicity parameters. 
Our results show that significant improvements in performance can be obtained employing a nonuniform grid.
\end{abstract}
%%%%% end %%%%%%%%%%%

%%%%% AMS/PACs/Keywords %%%%%%%%%%%
%\pac{}
%\ams{82D08, 82C08, 82B08}

%%%% maketitle %%%%%
\maketitle

\section{Introduction}

When a liquid is cooled or compressed towards structural arrest, its particles experience a slowing down of transport because they remain ``captured'' in transient ``cages'' formed by neighboring particles.
This phenomenon is known as the ``cage effect'' and is the underlying microscopic picture behind the glass transition~\cite{Goetze2009}.
The approach to structural arrest is connected with the appearance of several fascinating dynamical processes that manifest themselves in the low-frequency spectra or in the long-time behavior of time-correlation functions of the system~\cite{Goetze2009}.
Among these features, in particular, we recall the stretching of the response functions over time intervals extending over several orders of magnitude, a two-step relaxation process of the density-correlation function, and the aforementioned drastic slowing down of transport coefficients such as viscosity or diffusion.
All of these features have been observed both in experiments~\cite{Li1992,Li1992b,VanMegen1993,VanMegen1994,Wuttke1994,Yang1995,Cummins1997,Singh1998,Lunkenheimer1996,Schneider1999,Goetze2000} and molecular-dynamics simulations~\cite{Kob1994,Kob1995,Gallo1996,Sciortino1996,Sciortino1997,Horbach1998,Sciortino2001,Voigtmann2006,Das2008}.

In parallel to experiments and simulations, various theoretical frameworks provided different interpretations of the glass transition~\cite{Cavagna2009}. 
Among them, the mode-coupling theory (MCT) of the glass transition was originally developed to account for the cage effect in simple fluids by calculating explicitly the dynamics of density fluctuations. 
The theory is based on closed nonlinear integro-differential equations for a set of correlation functions where the coupling coefficients are determined by the equilibrium structural properties only.
In particular, the dependence on system parameters such as temperature and density is smooth and no assumptions on anomalous exponents, transitions or slow relaxations are built into the starting equations of the theory.
The success of MCT derives from the  many detailed predictions of striking features associated with the structural arrest~\cite{Goetze2009,Janssen2018,Bengtzelius1984,Franosch1997,Franosch1997b,Fuchs1998,Goetze1999,Goetze2002,Goetze2003}.

For numerical solutions of MCT equations, one has to discretize the wave-number dependence of the interesting functions such as the intermediate scattering functions and the structure factors.
Usually, this is done by taking a uniform grid of wave numbers.
The limiting factor in solving MCT equations is the evaluation of the MCT kernels (one for each wave number of the grid) which appear in the equations of motion of the correlation functions.
This evaluation has a computational cost that in principle scales as the third power of the grid size.

Nowadays, with modern CPUs, MCT equations for simple fluids in bulk can be solved in relatively short times.
However, some recent extensions of MCT require the introduction of matrix-valued correlation functions.
The consequence is that the computational effort required to solve the relative MCT dynamics becomes by orders of magnitude more demanding and, even more crucial, that the limits of computer allocation memory are easily exhausted.
This is the case, for example, of a generalization of MCT to multi-component fluids obtained through equations of motion that have to couple the different particle species~\cite{Bosse1987,Barrat1990,Nagele1999,Franosch2002}.
As a matter of fact, numerical solutions of multi-component MCT are up-to-date limited to 5 different species at most in the three-dimensional case~\cite{Weysser2010} and, due to the fact that the structure of the MCT equations is more complex, to 2 species only in the two-dimensional case~\cite{Hajnal2009,Weysser2011,Hajnal2011}.
Other examples, in which tensorial correlation functions appear, are advanced extensions of MCT apt to study the glassy behavior of molecular liquids~\cite{Schilling1997,Kammerer1998,Kammerer1998b,Goetze2000b,Theis2000,Letz2000,Winkler2000,Chong2002}, or active Brownian particles~\cite{Liluashvili2017}, or probe particles driven by a constant force through a colloidal glass~\cite{Gruber2016}.
Finally, matrix-valued correlation functions also appear in an extension of MCT describing simple fluids in confinement by introducing symmetry-adapted fluctuating density modes mirroring the broken translational symmetry~\cite{Lang2010,Franosch2012,Lang2012,Lang2014,Mandal2014,Mandal2017,Schrack2020}.
Intuitively, in the latter case the structure of MCT equations is more similar to the two-dimensional case than to the three-dimensional one and only recently a numerical solution for the full-time dependence of MCT equations in a confined system has been presented~\cite{Jung2020}.

In this paper we show that using a uniform grid for the wave numbers may not be the most clever choice in order to find an optimal compromise between the computational time and memory resources required to compute a solution of the MCT equations and the accuracy of such a solution.

Furthermore, relying on an equidistant grid for the MCT equations constitutes a serious limitation when one wants to resolve long-wavelength dynamics.
To overcome this limitation, only recently, a logarithmic grid has been used to compute the mean-square displacement 
or the velocity-autocorrelation function of Brownian particles in three dimensions~\cite{Mandal2019}.
However, a logarithmic grid has the disadvantage that a large number of grid points is necessary if one wants to cover the small wave numbers and, at the same time, to have enough grid points to resolve properly the first structure-factor peak driving the glass transition.
Therefore a new MCT integration scheme defined on a grid that can resolve well both the intermediate and the small wave numbers is highly desirable.
This is especially true in the two-dimensional case for which the structural relaxation in glassy dynamics is fundamentally different from the three-dimensional one~\cite{Flenner2015,vanderHoef1991} and the long-wavelength dynamics of glassy systems still has to be systematically addressed.

Here we propose two different schemes for nonuniform wave-number grids yielding a better representation of the structure-factor peak and compare them to the standard uniform discretization.
We exemplify the method both for the paradigmatic hard-sphere system ($d=3$) as well as for hard disks ($d=2$). 
Our results show that, with the new nonuniform grid, the number of grid points necessary to obtain an accurate solution can be lowered by a factor of two in comparison to the standard discretization, thus leading to a significant reduction of the computational effort.
We stress already now that our approach focuses only on the evaluation of the mode-coupling functional, leaving untouched the decimation scheme usually used for time integration.

The rest of the paper is organized as follows: 
Sec.~\ref{sec_MCT} presents the MCT equations in $d$ dimensions while Sec.~\ref{sec_integration} introduces the new nonuniform grid and the associated integration scheme of the MCT equations.
In Sec.~\ref{sec_results}, we report detailed results for the performance of the new grid and compare them with those of the standard discretization.
Finally, Sec.~\ref{sec_conclusions} summarizes and concludes the paper.

\section{Mode-coupling Theory} \label{sec_MCT}

In this section we briefly review the MCT equations for a mono-component simple liquid in bulk. 
For future reference, the dependence on the dimension $d$ of the bulk is accounted for explicitly until we specialize to $d=3$ or $d=2$.  

Density fluctuations $\rho(\vec{q},t) = \sum_{\alpha=1}^N \exp \left[ i \vec{q} \cdot \vec{r}_{\alpha}(t) \right] $ characterize the dynamics of a fluid consisting of $N$ structureless particles at position $\vec{r}_{\alpha}(t)$, $\alpha=1,\ldots,N$, at time $t$.
The simplest functions dealing statistically with the structure dynamics are the collective intermediate scattering functions
\begin{equation}\label{eq:densitycorr}
S(q,t) = \frac{1}{N} \langle \, \rho(\vec{q},t)^{\ast} \rho(\vec{q},0) \, \rangle \; ,
\end{equation}
which depend only on the magnitude $q = |\vec{q}|$ of the wave vector due to system isotropy.
Equations of motion of these quantities can be obtained within the Zwanzig-Mori formalism~\cite{McDonald1986,Forster1975,Boon1980}
\begin{equation}\label{eq:densitycorreqmotion0}
\ddot{S}(q,t)  +  \Omega^2(q) \, S(q,t) +  \int_0^t dt' \,\left[ M^{reg}(q,t-t') +  \Omega^2(q) \, m(q,t-t') \right]  \,  \dot{S}(q,t') = 0 \; , \qquad
\end{equation}
with initial conditions $S(q,t=0) = \langle | \rho(\vec{q},0)|^2 \rangle /N =: S(q)$ and $\dot{S}(q,t=0) = 0$. 
Here, $S(q)$ is the static structure factor and $\Omega (q) = q v/\sqrt{S (q)}$ is the characteristic frequency of the short-time dynamics, with $v$ denoting the thermal velocity.
For colloidal systems, Brownian dynamics is adopted, i.e. the regular relaxation kernel $M^{reg}(q,t)$ is Markovian, $M^{reg}(q,t) = \nu (q) \delta(t)$, and the friction constants $\nu (q)$ are assumed to be so large that the inertia terms can be neglected.
In this limit, Eq.~(\ref{eq:densitycorreqmotion0}) reduces to
\begin{equation}\label{eq:densitycorreqmotion}
\tau (q) \, \dot{S}(q,t) +  S(q,t) +  \int_0^t dt' \, m(q,t-t') \,  \dot{S}(q,t') = 0 \; ,
\end{equation}
with $\tau (q) = \nu(q)/\Omega^2(q)$.
The residual memory kernel $m(q,t)$ deals with the correlations of interparticle forces which are slowly fluctuating due to the slow relaxation of the system. 
In MCT, this kernel is given as a mode-coupling functional of the intermediate scattering functions, $m(q,t) = \mathcal{F} [S(t); q ]$, where $S(t)$ abbreviates the collection of the intermediate scattering functions for all possible wave numbers.
Using the factorization approximation~\cite{kawasaki1970}, the kernel can be expressed in terms of the density correlators and reads in the thermodynamic limit:
\begin{equation}\label{eq:MCTfunctionalthermolimit}
\mathcal{F} [S(t); q ] = \int \frac{\textrm{d}^d k}{(2\pi)^d} V(\vec{q};\vec{k},\vec{p}) \, S(k,t) \, S(p,t) \; ,
\end{equation}
where we use as abbreviation $\vec{p} = \vec{q}-\vec{k}$.
The vertices $V(\vec{q};\vec{k},\vec{p})$ are specified in terms of the structure factors~\cite{Bengtzelius1984,Sjogren1980}:
\begin{equation}\label{eq:MCTvertices}
V(\vec{q};\vec{k},\vec{p}) = \frac{n \, S(q) }{2q^4} \left[ \vec{q} \cdot \left( \vec{k} \, c(k) + \vec{p} \, c(p) \right) \right]^2  \; ,
\end{equation}
with $n$ the particle number density and $c(q)$ the direct correlation function related to $S(q)$ by the Ornstein-Zernike equation, which in Fourier space reads~\cite{McDonald1986}
\begin{equation} \label{eq_OZrelation}
S(q) = \frac{1}{1-n \, c(q)} \; .
\end{equation}

Introducing bipolar coordinates, Eq.~(\ref{eq:MCTfunctionalthermolimit}) reduces to a twofold integral~\cite{Bengtzelius1984}:
\begin{equation}\label{eq:MCTfunctionalthermolimittwofold}
\begin{split}
\mathcal{F} [S(t); q ] = \frac{n \, \Omega_{d-1} \, S(q)}{(4\pi)^d q^{d+2}} \int_0^{\infty} \textrm{d} k \, k \, S(k,t) \; \times \qquad \qquad \qquad \qquad  \qquad \qquad\\ 
 \qquad \qquad \int_{|q-k|}^{q+k} \textrm{d} p \, p \, S(p,t) \, \frac{\left[ (q^2+k^2-p^2) \, c(k) + (q^2+p^2-k^2) \, c(p)  \right]^2}{\left[ 4q^2k^2 - (q^2+k^2-p^2)^2 \right]^{(3-d)/2} } \; ,
\end{split}
\end{equation}
where $\Omega_d= 2 \pi^{d/2} / \Gamma(d/2)$ is the surface of the $d$-dimensional unit sphere ($\Omega_1=2$ and $\Omega_2=2\pi$).

\section{Numerical integration of the MCT equations} \label{sec_integration}

\subsection{Nonuniform grid} \label{sec_grid}
For the numerical computation of MCT kernels one has to discretize Eq.~(\ref{eq:MCTfunctionalthermolimittwofold}).
% Eq.s~(\ref{eq:MCTfunctionalthermolimittwofold}) and~(\ref{eq:MCTfunctionalthermolimittwofoldtagged}).
The standard way of doing that is to perform a Riemann-like summation defined on a uniform grid with a finite number of grid points $M$, which implies both a high- and a low-$q$ cutoff (respectively $q_{M-1}$ and $q_0$) with a discretization step $h=(q_{M-1}-q_0)/(M-1)$.
The exact value of the high-$q$ cutoff is not very important as long as $q_{M-1} \gtrsim 40/\sigma$ ($\sigma$ is the characteristic length scale of the exclusion region). 
However, the accuracy of the numerical solution considerably depends on the low-$q$ cutoff (see below).

\begin{figure}[!t]
	\centering
	\includegraphics[width=0.5\textwidth]{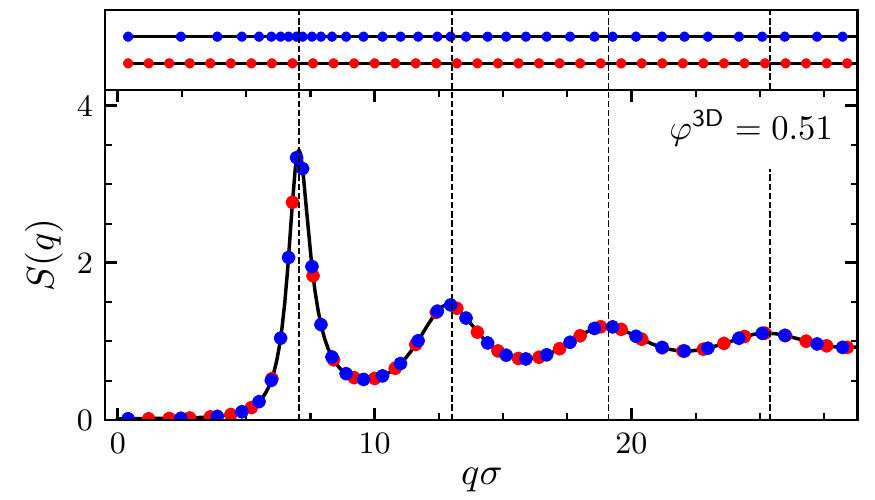} 
	\caption{Structure factor $S(q)$ as a function of wave vector $q$ for hard spheres at packing fraction $\varphi^{\mbox{\tiny{3D}}} = 0.51$  (solid black line). The discretization of the structure factor is represented by red dots for a uniform grid ($q_0 \sigma = 0.4, q_{M-1} \sigma = 34$, $h\sigma=0.8$ and $M=42$) and by blue dots for the nonuniform grid ($q_0 \sigma = 0.4, q_{M-1} \sigma = 34$ and $M=42$).
The upper panel is for a  better visual comparison between the two grids. \label{fig_grid}}
\end{figure}

In this paper we adopt a different strategy based on performing trapezoidal integration on a grid in which the grid points are nonuniformly distributed in the interval $[q_{0},q_{M-1}]$ according to a rule that determines the spacing between two consecutive grid points as inversely proportional to the quartic root of the absolute value of the second derivative of the structure factor.
We call this discretization nonuniform grid (\nhg).
We will show that, with the \nhg, the exact value of $q_{0}$ is not so crucial as long as $q_{0} \lesssim 1/\sigma$.
The main idea in favor of this grid is that it provides a better representation of the structure factor $S(q)$ (and consequently of the direct correlation function $c(q)$ and of the correlator $S(q,t)$) with more grid points around the values of $q$ where $S(q)$ has a maximum or a minimum and less grid points in the regions where $S(q)$ has a more linear behavior.
Fig.~\ref{fig_grid} shows the structure factor $S(q)$ for hard spheres (see Sec.~\ref{sec_results}) at packing fraction $\varphi^{\mbox{\scriptsize{3D}}} =  0.51$ and how the grid points of the standard grid and of the \nhg\ are distributed for the case of $q_0 \sigma= 0.3$, $q_{M-1} \sigma = 34$ and $M=42$.
When dealing with the \nhg , in the rest of the paper we fix $q_0 \sigma= 0.05$.
The reason for choosing the quartic root and not another power of the absolute value of the second derivative of the structure factor comes from the empirical observation that in this way we properly resolve the structure factor along the whole wave-number range.
For example, for grids of small size, taking the square root instead of the quartic route would lead to a grid that resolves very well the first peak but leave too few grid points for larger wave numbers.
Yet, the exponent leading to the best results depends on the grid size and in principle can be optimized through a numerical procedure.

 We also introduce a second kind of grid in which $M'<M-1$ grid points are logarithmically separated in the interval going from $q_0 = 10^{-4}/\sigma$ and $q_{M'} = 3/\sigma$ and the rest of the grid points are nonuniformly distributed in the interval $[q_{M'},q_{M-1}]$ according to the same rule based on the second derivative of the structure factor used for the \nhg .
We chose $M'=M/2$ if $M$ is even and $M'=(M-1)/2$ if $M$ is odd and we fix the grid points in the logarithmic part according to $q_i = x^{y+M'-i}$, where $i=0,\ldots,M'$, $x^y = q_{M'}$, and $x^{y+M'} = q_0$. 
Our choice $q_0 \sigma = 10^{-4}$ is justified by the observation that numerical instabilities appear for $q\sigma < 10^{-4}$ in three dimensions and for $q\sigma < 10^{-7}$ in two dimensions.
However, a smaller value of the low-$q$ cutoff could in principle be taken as long as one corrects the vertices, Eq.~(\ref{eq:MCTvertices}), by taking their Taylor expansion~\cite{Mandal2019}.

We refer to this discretization as logarithmic-nonuniform grid (\lognhg).
The \lognhg\ is somewhat similar to the grid used in Ref.~\cite{Gruber2016} where the authors combined a logarithmic grid with a uniformly distributed grid to study the dynamics of a probe particle driven by a constant force through a colloidal glass of hard spheres.

Because the denominator of the integrand in Eq.~(\ref{eq:MCTfunctionalthermolimittwofold})
%and (\ref{eq:MCTfunctionalthermolimittwofoldtagged})
, $\left[ 4q^2k^2 - (q^2+k^2-p^2)^2 \right]^{(3-d)/2}$, introduces an (integrable) divergence in the case of $d=2$, the integration scheme adopted for hard spheres ($d=3$) should be different from the one adopted for hard disks ($d=2$).
Then, from now on, we will distinguish between these two cases.

\subsection{Hard spheres ($d=3$)}

\begin{figure}[!t]
	\centering
	\includegraphics[width=1.0\textwidth]{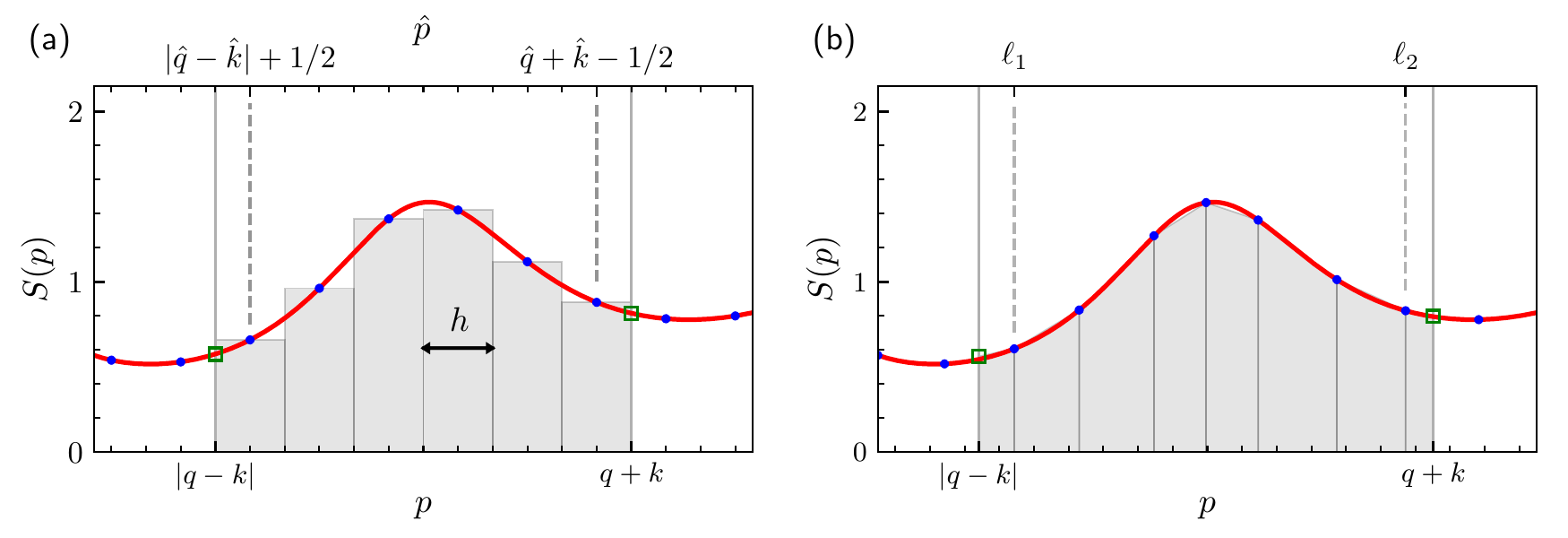} 
		\caption{Example of numerical integration of the inner integral in Eq.~(\ref{eq:MCTfunctionalthermolimittwofold}) for $d=3$.
The example integrand function is here the structure factor at $\varphi^{\mbox{\tiny{3D}}}= 0.51$.
	(a) Riemann integration on a uniform grid with $h \sigma=0.8$, $q=h\hat{q}$, $k=h\hat{k}$, $p=h\hat{p}$ and $\hat{q},\hat{k},\hat{p} =1/2,3/2,\ldots,M-1/2$.
	(b) Trapezoidal integration on the \nhg . Here, green points are obtained by linearly interpolating the integrand function between the two neighboring grid points belonging to the \nhg .
	 \label{fig_Riemann1}}
\end{figure}

The standard integration scheme approximates the mode-coupling functional by a Riemann sum~\cite{Franosch1997}.
%In three dimensions, this is done by setting $q_0 = h/2$ and rewriting Eq.s~(\ref{eq:MCTfunctionalthermolimittwofold}) and~(\ref{eq:MCTfunctionalthermolimittwofoldtagged}) as
In three dimensions, this is done by setting $q_0 = h/2$ and rewriting Eq.~(\ref{eq:MCTfunctionalthermolimittwofold}) as
\begin{equation}\label{eq:MCTfunctionaldiscrete3d}
\mathcal{F}_{\hat{q}} [S(t)] \! = \! \frac{n \, S_{\hat{q}} \, h^3}{32 \, \pi^2 \hat{q}^5 } \sum_{\hat{k} = 1/2,3/2,\ldots}^{M-1/2}  \hat{k} \, S_{\hat{k}}(t) \sum_{\hat{p}=|\hat{q}-\hat{k}|+1/2}^{\hat{q}+\hat{k}-1/2}  \hat{p} \, S_{\hat{p}}(t) \, \left[ (\hat{q}^2 \! + \! \hat{k}^2 \! - \! \hat{p}^2) \, c_{\hat{k}} + (\hat{q}^2 \! + \! \hat{p}^2 \! - \! \hat{k}^2) \, c_{\hat{p}}  \right]^2 \; ,
\end{equation}
where we used the half-integer indices $\hat{q},\hat{k},\hat{p} =1/2,3/2,\ldots,M-1/2$ and $q=h\hat{q}$, $S_{\hat{q}} = S(h\hat{q})$, $c_{\hat{q}} = c(h\hat{q})$ and $S_{\hat{q}}(t) = S(h\hat{q},t)$.
Note that in the previous equation, the inner summation is restricted to $|\hat{q}-\hat{k}|+1/2 \leq \hat{p} \leq \hat{q}+\hat{k}-1/2$, while the integration domain of the inner integral in Eq.~(\ref{eq:MCTfunctionalthermolimittwofold}) extends from $|q-k|$ to $q+k$.
Here, the choice $q_0 = h/2$ turns out to be the proper one for the integration scheme just described.
In fact, dividing the interval $[|q-k|,q+k]$ by $h$, one obtains an integer number of subintervals and the grid points are correctly defined in the center of these subintervals (see Fig.~\ref{fig_Riemann1}a).
\medskip

Using the \nhg\ or the \lognhg , the spacing between consecutive grid points is not constant.
Furthermore, when considering the inner integral in Eq.~(\ref{eq:MCTfunctionalthermolimittwofold}), also the spacing between the extrema of the integration domain and the neighboring grid points is not fixed.
As a consequence, the Riemann sum is not suitable anymore to approximate the integral.
We then resort to trapezoidal integration:
\begin{equation}\label{eq:MCTfunctionaldiscretetrap3d}
\mathcal{F}_{i} [S(t)] = \frac{n \, S_i}{32 \, \pi^2 q_i^5 } \sum_{j = 0}^{M-2} \frac{q_j \, S_j (t) \, A_{i,j} + q_{j+1} \, S_{j+1} (t) \, A_{i,j+1}}{2} \, \left( q_{j+1} - q_{j} \right)  \; ,
\end{equation}
where $i=0,\ldots,M-1$ is an integer index and we used the simplified notation $S_i = S(q_i)$, $S_i(t) = S(q_i,t)$ with $q_i$ the value of the $i$-th grid wave number. The inner integral of Eq.~(\ref{eq:MCTfunctionalthermolimittwofold}) is here approximated by the function $A_{i,j}$.
If we call $\ell_1$ the index $\ell$ of the first grid point for which  $q_{\ell} > |q_i - q_j|$, and $\ell_2$ the index $\ell$ of the last grid point for which $q_{\ell} < (q_i + q_j)$ (see Fig.~\ref{fig_Riemann1}b), then we can write:
\begin{equation}
\begin{array}{cl}
A_{i,j} =   & \displaystyle  \frac{2B_{\ell_1} + \displaystyle \frac{q_{\ell_1} -|q_i-q_j|}{q_{\ell_1}-q_{\ell_1-1}}(B_{\ell_1-1}-B_{\ell_1}) }{2} \left( q_{\ell_1} - |q_i-q_j| \right)   \\
& \, \\
& + \displaystyle \sum_{\ell=\ell_1}^{\ell_2-1} \displaystyle \frac{B_{\ell} + B_{\ell+1}}{2} \left( q_{\ell+1} - q_{\ell} \right) +   
\frac{2B_{\ell_2} + \displaystyle \frac{q_i+q_j-q_{\ell_2}}{q_{\ell_2+1}-q_{\ell_2}}(B_{\ell_2+1}-B_{\ell_2}) }{2} \left(q_i+q_j - q_{\ell_2} \right)  \; ,
\end{array}
\end{equation}
with
\begin{equation}
B_{\ell} =  q_{\ell} \, S_{\ell}(t)  \, \left[ (q_i^2+q_j^2-q_{\ell}^2) \, c_j + (q_i^2+p_{\ell}^2-q_j^2) \, c_{\ell}  \right]^2 \; ,
\end{equation}
and where we have approximated the value of the functional $B$ for the wave number equal to $|q-k|$ by linearly interpolating between its values taken at the grid points with index $\ell_1 -1$ and $\ell_1$.
Similarly, for the wave number equal to $q+k$ the functional $B$ is approximated by linearly interpolating between the grid points with index $\ell_2$ and $\ell_2+1$ (see Fig.~\ref{fig_Riemann1}b).

The integration scheme proposed here for $d=3$ can in principle be used for any dimension $d\geq 3$ and thus can be helpful to explore the MCT scenario also in higher dimensions~\cite{Schmid2010,Ikeda2010,Schilling2011}.
Furthermore, we note that the evaluation of the MCT functional as given by Eq.~(\ref{eq:MCTfunctionaldiscrete3d}) has a computational cost that scales as $M^2$ and this functional must be evaluated at each grid point, thus bringing a total computational cost $\mathcal{O}(M^3)$.
However, for $d=3$ (as for all odd $d \geq 3$) it is possible to reduce the total computational to $\mathcal{O}(M^2)$ using the Bengtzelius factorization~\cite{Bengtzelius1984,sperl2000}, i.e.  by expanding the square in the second sum of Eq.~(\ref{eq:MCTfunctionaldiscrete3d}) and rearranging the terms in such a way that the second summation can be calculated recursively.
Here we do not enter in details but we stress the fact that the same reduction of computational cost can be achieved in the \nhg\ or \lognhg\ framework by starting again from Eq.~(\ref{eq:MCTfunctionalthermolimittwofold}) and following the same idea.

\subsection{Hard disks ($d=2$)}\label{sec_hard_disks_integration}

\begin{figure}[!t]
	\centering
	\includegraphics[width=1.0\textwidth]{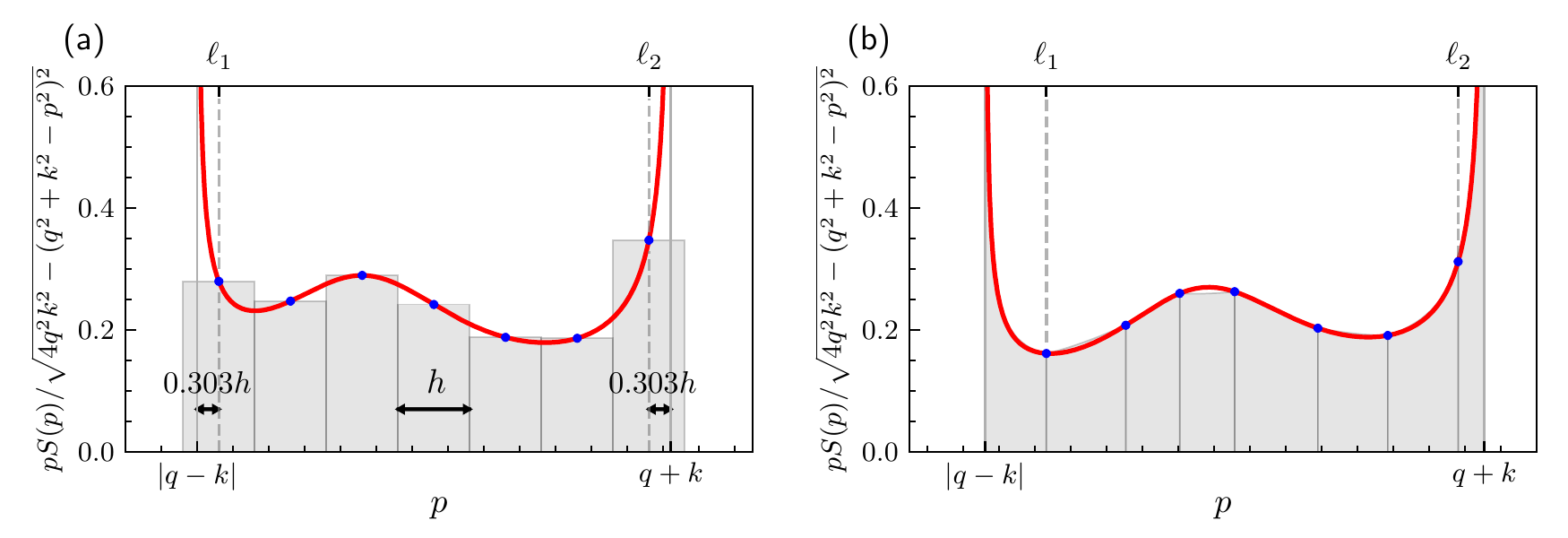} 
	\caption{Example of numerical integration of the second integral in Eq.~(\ref{eq:MCTfunctionalthermolimittwofold}) for $d=2$.
The example integrand function is here the function $pS(p)/\sqrt{4q^2k^2-(q^2+k^2-p^2)^2}$ at $\varphi^{\mbox{\tiny{2D}}} = 0.69$, $q\sigma \simeq 12.6$, $k\sigma \simeq 2.8$.
	(a) Riemann integration on a uniform grid with $h \sigma=0.8$, $q_i=(i+0.303)h$, with $i=0,\dots,M-1$.
	(b) Semi-analytical integration on the \nhg\ as described in section~\ref{sec_hard_disks_integration}.
	Here, since $\int dp \, p \, S(p)/\sqrt{4q^2k^2-(q^2+k^2-p^2)^2} = \int dx \, S(p) / \sqrt{1-x^2} $, with $x=(q^2+k^2-p^2)/2q k$, we have only assumed that the function $S(p)$ is linear between two neighboring grid points belonging to the \nhg .
	 \label{fig_Riemann2}}
\end{figure}

In two dimensions, the integrand of the second integral of Eq.~(\ref{eq:MCTfunctionalthermolimittwofold}) is singular for $p=|q-k|$ and $p=q+k$.
The traditional way of integrating MCT equations remains Riemann-like~\cite{Bayer2007} but with the precaution that the grid points must be defined by $q_i=(i+0.303)h$, with $i=0,\dots,M-1$~\cite{Bayer2007}.
This distinction from the grid used in three dimensions is necessary in order to obtain the best discrete description of the Jacobian $[4q^2k^2 - (q^2+k^2-p^2)^2]^{-1/2}$ of the transformation to bipolar coordinates.
In this way the asymptotes of the inner integrand in Eq.~(\ref{eq:MCTfunctionalthermolimittwofold}) are always at a distance equal to $0.303$h from the closest grid point inside the integration domain (see Fig.~\ref{fig_Riemann2}a).
Also, this grid definition fixes the low-$q$ cutoff at $0.303h$.

Again, if we call $\ell_1$ the index $\ell$ of the first grid point for which  $q_{\ell} > |q_i - q_j|$, and $\ell_2$ the index $\ell$ of the last grid point for which $q_{\ell} < (q_i + q_j)$ (see Fig.~\ref{fig_Riemann2}a), Eq.~(\ref{eq:MCTfunctionalthermolimittwofold}) can be discretized as
\begin{equation}\label{eq:MCTfunctionaldiscrete2d}
\mathcal{F}_{i} [S(t)] = \frac{n \, S_i \, h^2}{8 \pi^2 q_i^4} \sum_{j=0}^{M-1}  q_j \, S_j(t) \sum_{\ell=\ell_1}^{\ell_2}  q_{\ell} \, S_{\ell}(t) \, \frac{\left[ (q_i^2+q_j^2-q_{\ell}^2) \, c_j + (q_i^2+q_{\ell}^2-q_j^2) \, c_{\ell}  \right]^2}{\left[ 4 \, q_i^2q_j^2 - (q_i^2+q_j^2-q_{\ell}^2)^2 \right]^{1/2} } \; ,
\end{equation}
where again we simplify notation by writing $S_i = S(q_i)$, $S_i(t) = S(q_i,t)$ and $c_i = c(q_i)$.
\medskip

Also in two dimensions, if we want to use a nonuniform grid, we have to abandon Riemann sums as a mean to approximate Eq.~(\ref{eq:MCTfunctionalthermolimittwofold}).
Here, the integration scheme that we adopt to compute the kernel is trapezoidal for the outer integral, $\int \textrm{d} k \ldots$, of Eq.~(\ref{eq:MCTfunctionalthermolimittwofold}) 
\begin{equation}\label{eq:MCTfunctionaldiscretetrap2d}
\mathcal{F}_{i} [S(t)] = \frac{n \, S_i}{8 \, \pi^2 q_i^4 } \sum_{j = 0}^{M-2} \frac{q_j \, S_j(t)  \, A_{i,j} + q_{j+1} \, S_{j+1} (t) \, A_{i,j+1}}{2} \, \left( q_{j+1} - q_{j} \right)  \; ,
\end{equation}
and \textbf{semi-analytical} on the inner integral, $A_{i,j} \approx \int \textrm{d} p \ldots$ .
To write an expression for the functional $A_{i,j}$, we first notice that:
\begin{eqnarray} \label{eq_appendixA1}
A_{i,j} \approx \int_{|q_i-q_j|}^{q_i+q_j} \textrm{d} p \, p \, S(p,t)
\frac{\left[ (q_i^2+q_j^2-p^2)\, c_j + (q_i^2+p^2-q_j^2)\, c(p)  \right]^2}{\sqrt{4q_i^2q_j^2 - (q_i^2+q_j^2-p^2)^2}} = \qquad \qquad \qquad \qquad \nonumber \\
 = 2 q_i^2 q_j^2 \int_{-1}^1 \textrm{d} x \, S(p,t) \, \frac{\left[ x \left(  c_j-c(p) \right)  + \frac{q_i}{q_j}c(p) \right]^2}{\sqrt{1-x^2}} \; , \qquad
\end{eqnarray}
upon a change of variables  $p \mapsto x=(q_i^2+q_j^2-p^2)/2q_i q_j$ . Here one has to keep in mind that by the substitution  $S(p,t)$ and $c(p)$ now depend also on $x$.
Then, we split the interval $[-1,1]$ in $\ell_2-\ell_1+2$ subintervals where $\ell_2-\ell_1+1$ is the number of grid points contained in the interval $[|q_i-q_j|,q_i+q_j]$ and we express the integral in the second line of the previous equation as a sum of integrals on these subintervals
\begin{equation} \label{eq_appendixA2}
 \int_{-1}^1 \textrm{d} x \, S(p,t) \, \frac{\left[ x \left(  c_j-c(p) \right) + \frac{q_i}{q_j}c(p) \right]^2}{\sqrt{1-x^2}} 
 = \sum_{\ell=\ell_1-1}^{\ell_2} \int_{x_{\ell+1}}^{x_{\ell}} \textrm{d} x \, S(p,t) \, \frac{\left[ x \left(  c_j-c(p) \right) + \frac{q_i}{q_j}c(p) \right]^2}{\sqrt{1-x^2}} 
 \; ,
\end{equation}
where $x_{\ell_1-1}=-1$ (corresponding to $p=q_i+q_j$), $x_{\ell_2+1}=1$ (corresponding to $p=|q_i-q_j|$) and $x_{\ell}=(q_i^2+q_j^2-p_{\ell}^2)/2qk$ with the index $\ell=\ell_1,\ldots,\ell_2$ labeling the grid points $q_{\ell}$  for which the inequality $|q_i-q_j|<q_{\ell}<q_i+q_j$ holds.
Finally, we rewrite the integrals in the right term of Eq.~(\ref{eq_appendixA2}) 
\begin{equation} \label{eq_appendixA3}
\int_{x_{\ell+1}}^{x_{\ell}} \textrm{d} x \, S(p,t) \, \frac{\left[ x  \left(  c_j-c(p) \right) + \frac{q_i}{q_j} c(p) \right]^2}{\sqrt{1-x^2}}
= \sum_{\alpha=0}^2 \int_{x_{\ell+1}}^{x_{\ell}} \textrm{d} x \, g_{\alpha}(x; q_i,q_j) \, \frac{x^{\alpha}}{\sqrt{1-x^2}}
 \; ,
\end{equation}
with $g_0(x; q_i,q_j) = [q_i c(p)/q_j]^2 S(p,t)$, $g_1(x; q_i,q_j) = 2 [c_j-c(p)][q_i c(p)/q_j]S(p,t)$ and $g_2(x; q_i,q_j) = [c_j-c(p)]^2 S(p,t)$ (remember $x=(q_i^2+q_j^2-p^2)/2q_i q_j$), and we solve them by approximating the functions $g_\alpha$ as linear functions in $x$ in the small interval $[x_{\ell+1},x_{\ell}]$ and making use of the known integrals
\begin{eqnarray} \label{eq_appendixA4}
\int \textrm{d} x \frac{1}{\sqrt{1-x^2}} & = & \arcsin(x) \, ,
 \\
\int \textrm{d} x \frac{x}{\sqrt{1-x^2}} & = & - \sqrt{1-x^2} \, , 
 \\
\int \textrm{d} x \frac{x^2}{\sqrt{1-x^2}} & = & \frac{1}{2}\arcsin(x) - \frac{x}{2}\sqrt{1-x^2}  \, , 
 \\ 
 \int \textrm{d} x \frac{x^3}{\sqrt{1-x^2}} & = & -\frac{1}{3}\sqrt{1-x^2} (2+x^2) \, .
\end{eqnarray}

\section{Results} \label{sec_results}

In this section we apply MCT to a system of hard particles with diameter $\sigma$.
The only system control parameter that we vary is the packing fraction $\varphi^{\mbox{\scriptsize{3D}}} = n \pi \sigma^3/6$ for hard spheres and $\varphi^{\mbox{\scriptsize{2D}}} = n \pi \sigma^2/4$ for hard disks.

\subsection{Hard spheres ($d=3$)}

As input we take the structure factors $S(q)$ as obtained from the Ornstein-Zernike equation (\ref{eq_OZrelation}) and direct correlation function $c(q)$ given in the Wertheim and Thiele representation~\cite{Thiele1963,Wertheim1963} within the Percus-Yevick approximation~\cite{Percus1958}
\begin{eqnarray} \label{eq_directcorrfunction3d}
c(q)  =  - \frac{4 \pi}{q^6} \, \Big\lbrace \alpha q^3 \left[ \sin(q) - q\cos(q) \right] 
 + \beta q^2 \left[ 2q \sin(q) - (q^2-2) \cos(q) -2 \right] + \qquad \qquad \nonumber \\
 + \gamma \left[ (4q^3-24q)\sin(q) - (q^4-12q^2+24)\cos(q) +24 \right] \Big\rbrace \; ,
\end{eqnarray}
where
\begin{equation}
\alpha = \frac{(1+2\varphi)^2}{(1-\varphi)^4} \; , \qquad
\beta  = -6 \, \varphi \frac{(1+\varphi/2)^2}{(1-\varphi)^4 } \; , \qquad
\gamma = \frac{\alpha}{2} \varphi \; .
\end{equation}

To test the performances of the proposed integration scheme once applied to the \nhg\ described in Section~\ref{sec_integration}, in comparison to the standard integration scheme with a uniform grid, we compute the critical packing fraction in $d=3$, $\varphi^{\mbox{\scriptsize{3D}}}_c$ through a standard bisection algorithm, with initial guesses $\varphi^{\mbox{\scriptsize{3D}}}=0.4$ and $\varphi^{\mbox{\scriptsize{3D}}}=0.6$.
To find the critical point we evaluate the nonergodicity parameters
\begin{equation} \label{eq_NEP}
F(q) = \lim_{t\rightarrow \infty} S(q,t) \; ,
\end{equation}
considering that they change discontinuously from zero when $\varphi^{\mbox{\scriptsize{3D}}} < \varphi^{\mbox{\scriptsize{3D}}}_c$ to a positive nonzero value when $\varphi^{\mbox{\scriptsize{3D}}} \geq \varphi^{\mbox{\scriptsize{3D}}}_c$ and that they fulfill the system of equations~\cite{Bengtzelius1984}:
\begin{equation} \label{eq_NEPqueation}
\frac{F(q)}{S(q)-F(q)} = \mathcal{F} [S(t=\infty); q ] \; . 
\end{equation}

\begin{table*}[!t]
\centering
\scriptsize{
\begin{tabular}{|c|ccc|ccc|ccc|}
\hline 
 & \multicolumn{3}{c|}{{\small uniform grid}} & \multicolumn{3}{c|}{{\small \nhg}} & \multicolumn{3}{c|}{{\small \lognhg}} \\
\hline
& & & & & & & & & \\[-2ex]
{\footnotesize $M$} & {\footnotesize $\varphi^{\mbox{\scriptsize{3D}}}_c$} & {\footnotesize $\% \mbox{diff.}$} &  {\scriptsize time [sec]} & {\footnotesize $\varphi^{\mbox{\scriptsize{3D}}}_c$} & {\footnotesize $\% \mbox{diff.}$} &  {\scriptsize time [sec]} & {\footnotesize $\varphi^{\mbox{\scriptsize{3D}}}_c$} & {\footnotesize $\% \mbox{diff.}$} &  {\scriptsize time [sec]} \\
\hline 
& & & & & & & & & \\ [-2ex]
{\footnotesize $ 25$} & $0.57414105$ &$+11.30$ & $      4.9$     & $0.49273144$ & $-4.44$ & $     15.1$     & $0.45328619$ & $-12.08$ & $    49.0$ \\ 
{\footnotesize $ 50$} & $0.50440301$ & $-2.22$ & $     41.3$     & $0.50844091$ & $-1.40$ & $     95.9$     & $0.50606337$ & $-1.79$ & $    283.1$ \\ 
{\footnotesize $100$} & $0.51698310$ & $+0.22$ & $    333.8$     & $0.51359349$ & $-0.40$ & $    990.4$     & $0.51071595$ & $-0.89$ & $   2073.9$ \\ 
{\footnotesize $200$} & $0.51585338$ & $-0.01$ & $   4359.4$     & $0.51517387$ & $-0.09$ & $   7794.9$     & $0.51403430$ & $-0.25$ & $  17466.1$ \\ 
{\footnotesize $400$} & $0.51586705$ & $ 0.00$ & $  32165.3$     & $0.51565060$ & $ 0.00$ & $  61263.7$     & $0.51531808$ & $ 0.00$ & $ 157474.5$ \\ 
\hline 
\end{tabular}
}
\caption{\label{tab_table3d}Critical packing fraction for hard spheres ($d=3$) and computational time to evaluate it on a standard uniform grid and the nonuniform grid (\nhg) by varying the number of points $M$.
$\% \mbox{diff.}$ represents the percentage difference with respect to the exact critical packing fraction estimated by the value $\varphi^{\mbox{\tiny{3D}}}_c$ obtained with the finest grid ($M=400$).
For both grids $q_{M-1} \sigma \simeq 80$ with $q_0$ determined according to the grid kind (see text in section~\ref{sec_grid}). The computational time is the time required to reach the critical packing fraction to a precision of $10^{-8}$ on a single core 3.10 GHz CPU.}
\end{table*}

Table~\ref{tab_table3d} reports the results obtained when varying the size of the grid $M$ and the grid points are displaced in between $q_0$ and $q_{M-1}$ according to the grid type (see section~\ref{sec_grid}) and $q_{M-1} \sigma \simeq 80$.
At the smallest values of the grid size, namely $M=25$ and $M=50$, the \nhg\ is the grid that returns the best results in the estimation of the critical packing fraction, with a systematic improvement by about a factor two with respect to the results obtained with the uniform grid.
However, the MCT kernel integration scheme used with the \nhg\ is more complex and this results in a total computational time necessary to find the critical packing fraction which is in general a factor $2$ or $3$ higher with respect to the time required by the standard method at the same value of $M$.
Unfortunately the \lognhg\ grid  displays the worst performances both from the point of view of the precision in the computation of the critical packing fraction and of the total computational time required to do that.
Interesting enough, a notable exception is the case of $M=50$ for which the percentage difference with respect to critical packing fraction obtained at $M=400$ is comparable to that obtained with the \nhg .
Nevertheless, it would be easy to improve the performances of the \lognhg\ by increasing the fraction between the number of grid points placed in the nonuniform part of the grid and the number of points distributed in the logarithmic part of the grid.

\begin{figure}[!t]
	\centering
	\includegraphics[width=1.0\textwidth]{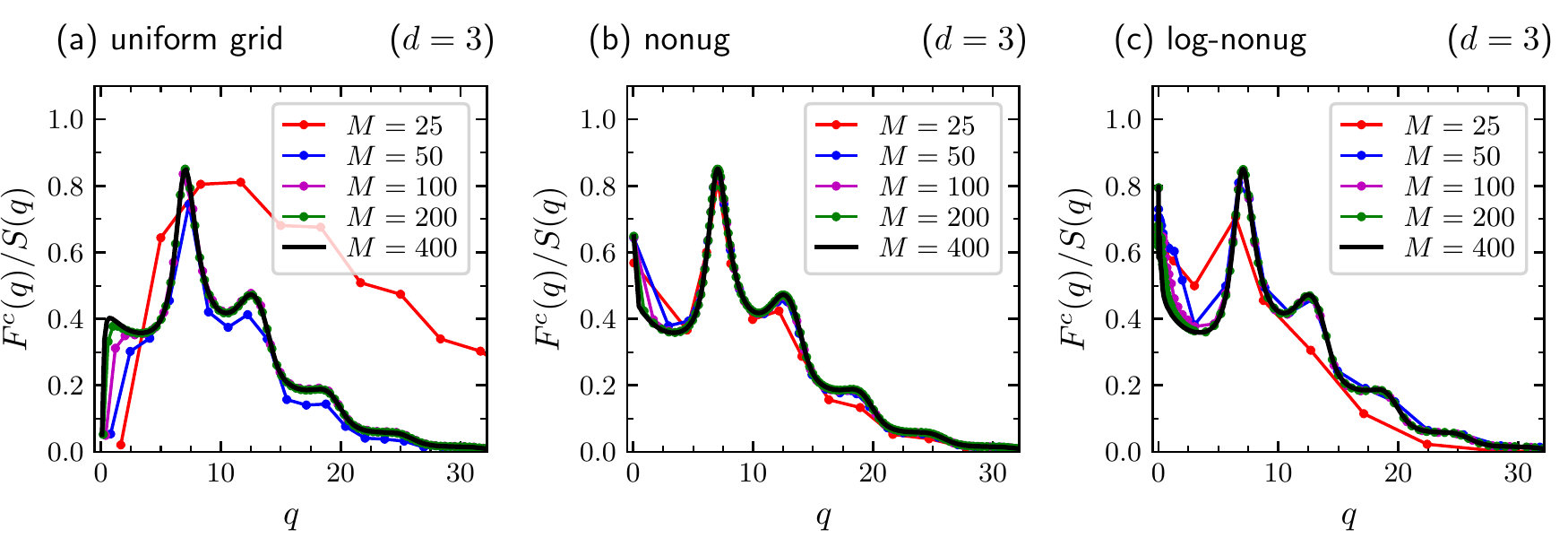} 
	\caption{Normalized nonergodicity parameters at the critical packing fraction $F^c(q)/S(q)$ as a function of wave vector $q$ for hard spheres. For each value of $M$ the critical packing fraction is the one reported in Table~\ref{tab_table3d}. \label{fig_Fc3d}}
\end{figure}
The observation based on Table~\ref{tab_table3d} are confirmed by the plots of the nonergodicity parameters as a function of the wave number $q$ (see Fig.~\ref{fig_Fc3d}).
As expected, when increasing $M$, the nonergodicity parameters collapse towards the same results.
Yet, it is interesting to notice that for $M=25$ the nonergodicity parameters obtained with the uniform grid and, to a less extent, those obtained with the \lognhg , are very far from the correct curve while, with the \nhg , they already give a good representation of the exact function.
Similar considerations apply to the case of $M=50$.
Another interesting aspect emerging from Fig.~\ref{fig_Fc3d} is the behavior of the nonergodicity parameters for $q\sigma \lesssim 1$.
While $F^c(q)/S(q)$ drops down to values close to zero when integrating with the uniform grid, integration on the \nhg\ leads to a value that in the limit of $q \rightarrow 0$ is about $0.6$.
This is more consistent with the value of about $0.45$ expected from the analytic expansions of Eq.~(\ref{eq:MCTfunctionalthermolimit})~\cite{Bayer2007}.

\subsection{Hard disks ($d=2$)}

For hard disks the direct correlation functions $c(q)$ are reconstructed within the framework of fundamental measure theory~\cite{Rosenfeld1989,Roth2010,Thorneywork2018}:
\begin{equation} \label{eq_directcorrfunctionFMT}
\begin{split}
c(q)  =  \frac{\pi}{6(1-\varphi)^3 q^2} \, \bigg{\lbrace} -\frac{5}{4} (1-\varphi)^2  q^2 \sigma^2 \mbox{J}_0(q\sigma/2)^2   
 + \\
 \left[ 4 \left( (\varphi-20)\varphi +7\right)  + \frac{5}{4} (1-\varphi)^2  q^2 \sigma^2 \right] \mbox{J}_1(q\sigma/2)^2 +  \\ 
 2(\varphi-13)(1-\varphi) q \sigma \mbox{J}_1(q\sigma/2) \mbox{J}_0(q\sigma/2)  \bigg{\rbrace}  \; , \qquad
 \end{split}
\end{equation}
where $\mbox{J}_0(x)$ and $\mbox{J}_1(x)$ are respectively the order zero and order one Bessel functions of the first kind.
The structure factors $S_q$ are then obtained from the Ornstein-Zernike equation (\ref{eq_OZrelation}).

\begin{figure}[!t]
	\centering
	\includegraphics[width=1.0\textwidth]{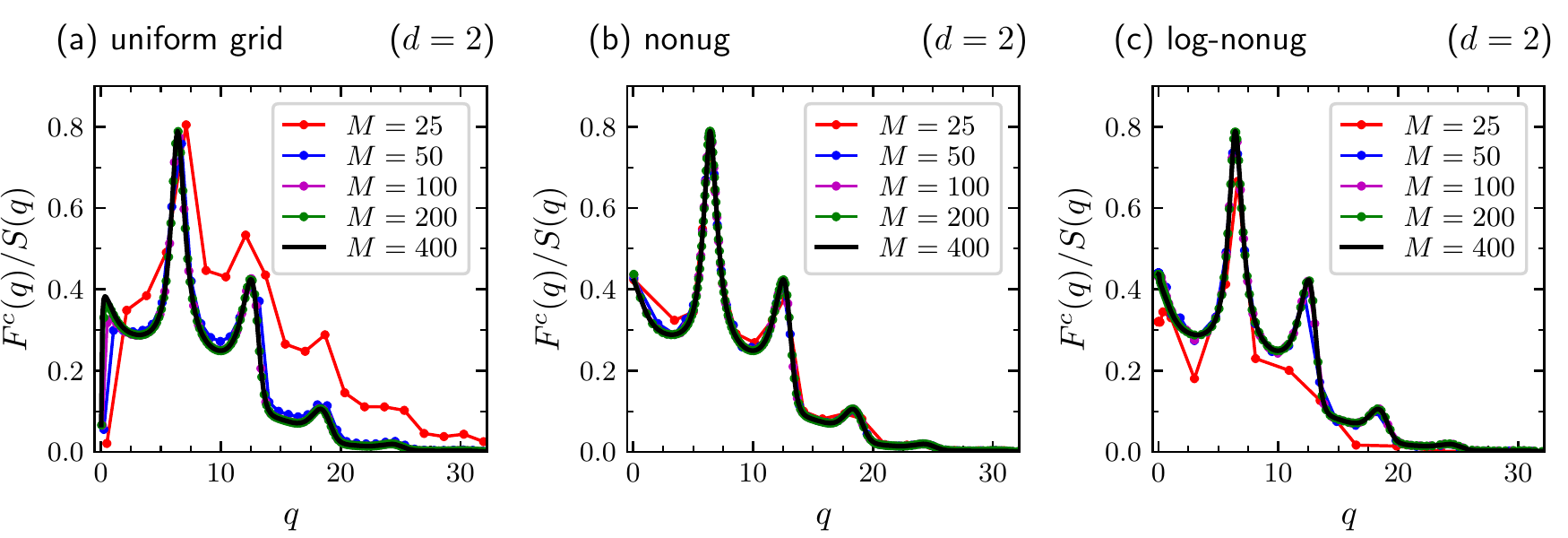} 
	\caption{Normalized nonergodicity parameters at the critical packing fraction $F^c(q)/S(q)$ as a function of wave vector $q$ for hard disks. For each value of $M$ the critical packing fraction is the one reported in Table~\ref{tab_table2d}. \label{fig_Fc2d}}
\end{figure}

Table~\ref{tab_table2d} is the equivalent of Table~\ref{tab_table3d} in two dimensions and reports the results obtained when varying the number of grid points $M$.
In this case we used $q_{M-1} \sigma \simeq 40$.
As expected, the critical packing fractions obtained from numerical integration on the three different grids, converge towards the same value when increasing $M$.
When decreasing $M$, MCT integration on the \nhg\ with the integration scheme proposed in section~\ref{sec_hard_disks_integration} appears to be the most stable and  also the \lognhg\ performs well with a percentage difference with respect to the reference value not exceeding $1$\% if $M\geq50$.
Interestingly enough, the computational time necessary to compute the critical packing fraction for the \lognhg\ is comparable to the time required with the standard method and smaller than the computational time required by computation on the \nhg .
This observation can be explained considering that with our method, given the higher complexity of the integration scheme, the computational cost of a single kernel evaluation is higher when compared to the standard method, but the number of iterations necessary to solve Eq.~(\ref{eq_NEPqueation}) and to find the nonergodicity parameters is lower in the case 
of the \lognhg.
Again, these findings are confirmed by the plots of the normalized nonergodicity parameters at the critical packing fraction reported in Fig.~\ref{fig_Fc2d}.
As for the case $d=3$, for the low values of $M$ ($M=25$ and $M=50$) the nonergodicity parameters obtained with the \nhg\ and the \lognhg\ represent the exact curve better.
Furthermore, also in the present case of $d=2$, the behavior at low values of $q$ is very different depending on the grid.
With the uniform grid $F^c(q)/S(q)$ drops down to values close to zero when $q \rightarrow 0$ while it reaches a value of about $0.45$ with \nhg\ and the \lognhg , which is more consistent  with the value of about $0.5$ expected from the analytic expansions of Eq.~(\ref{eq:MCTfunctionalthermolimit})~\cite{Bayer2007}.
All these observations make the use of the \lognhg\ particularly interesting and these results look even more promising if one takes into consideration that half of the grid points are placed on the logarithmic-grid part in between $q_0 = 10^{-5}/\sigma$ and $q_{M/2} = 3/\sigma$ and they should not contribute too much to the determination of the critical packing fraction.

\begin{table*}[!t]
\centering
\scriptsize{
\begin{tabular}{|c|ccc|ccc|ccc|}
\hline 
 & \multicolumn{3}{c|}{{\small uniform grid}} & \multicolumn{3}{c|}{{\small \nhg}} & \multicolumn{3}{c|}{{\small \lognhg}} \\ 
\hline 
& & & & & & & & & \\ [-2ex]
{\footnotesize $M$} & {\footnotesize $\varphi^{\mbox{\scriptsize{2D}}}_c$} & {\footnotesize $\% \mbox{diff.}$} &  {\scriptsize time [sec]} & $\varphi^{\mbox{\scriptsize{2D}}}_c$ & {\footnotesize $\% \mbox{diff.}$} &  {\scriptsize time [sec]} & $\varphi^{\mbox{\scriptsize{2D}}}_c$ & {\footnotesize $\% \mbox{diff.}$} &  {\scriptsize time [sec]} \\
\hline
& & & & & & & & & \\ [-2ex]
{\footnotesize $ 25$} & {\scriptsize $0.74808566$} & {\scriptsize $+8.31$} & {\scriptsize $      7.8$}     & {\scriptsize $0.68214627$} & {\scriptsize $-1.22$} & {\scriptsize $     19.3$}     & {\scriptsize $0.67858137$} & {\scriptsize $-1.73$} & {\scriptsize $     21.4$} \\
{\footnotesize $ 50$} & $0.70160688$ & $+1.58$ & $     61.4$     & $0.68810027$ & $-0.36$ & $    350.1$     & $0.68447831$ & $-0.87$ & $    113.8$ \\ 
{\footnotesize $100$} & $0.69130231$ & $+0.09$ & $    521.5$     & $0.68999410$ & $-0.09$ & $   3928.9$     & $0.68913991$ & $-0.20$ & $    336.6$ \\ 
{\footnotesize $200$} & $0.69076595$ & $+0.01$ & $   7651.3$     & $0.69045644$ & $-0.02$ & $  43099.4$     & $0.69018070$ & $-0.05$ & $   7407.2$ \\ 
{\footnotesize $400$} & $0.69068515$ & $ 0.00$ & $  31818.8$     & $0.69058952$ & $ 0.00$ & $ 293932.5$     & $0.69051168$ & $ 0.00$ & $  66143.2$ \\
\hline 
\end{tabular}
}
\caption{\label{tab_table2d}Critical packing fraction for hard disks ($d=2$) and computational time to evaluate it through a ``classic'' uniform grid and the nonuniform grid (\nhg) by varying the number of points $M$.
$\% \mbox{diff.}$ represents the percentage difference with respect to the exact critical packing fraction estimated by the value $\varphi^{\mbox{\tiny{2D}}}_c$ obtained with the finest grid ($M=400$).
For both the grids $q_M \sigma \simeq 40$ with $q_0$ determined according to the grid kind (see text). The computational time is the time required to reach the critical packing fraction to a precision of $10^{-8}$ on a single core 3.10 GHz CPU.}
\end{table*}

\section{Conclusions} \label{sec_conclusions}

We have reconsidered the standard integration scheme for MCT equations for the density-fluctuation dynamics of hard spheres and hard disks in bulk and we have tested a different integration scheme able to deal with a nonuniform grid in wave number space.
Here we investigated two different nonuniform grids: $i$) a grid whose points are nonuniformly distributed in the interval according to a rule that determines the spacing between two consecutive grid points as inversely proportional to the quartic root of the absolute value of the second derivative of the structure factor (\nhg ) and $ii$) a grid in which half of the grid points are logarithmically separated, thus covering the low-wave numbers, and half of the grid points are nonuniformly distributed following the same rule of the \nhg\ (\lognhg ).

The main merit of the grids we propose here is that, depending on the number of grid points, they may provide a better representation of the structure factor and consequently of the direct correlation function and of the density correlators.
Furthermore, the \lognhg\ allows studying numerically the long-wavelength regime $q \rightarrow 0$.
This could be particularly interesting to study the long-wavelength properties of the structural glass transition in two dimensions, a topic that is recently receiving increasing attention~\cite{Illing2017,Shiba2019}.
The proposed integration scheme resorts completely on trapezoidal integration in the case of hard spheres, while for hard disks we introduced a mix of trapezoidal and analytical integration that allows us to deal successfully with the singularities appearing in the MCT kernel for $d=2$.

When computing the critical packing fraction, the results show that, in comparison to standard integration, better results can be achieved using trapezoidal integration on the \nhg .
Depending on the level of accuracy that one wants to achieve, these results suggest that using the \nhg\ it is possible to decrease the number of grid points by a factor two, which, taking into account the increased complexity of the algorithm, would lead to a decrease of computational time by about a factor 3 or 4.
On the other hand, the integration on the \lognhg\ in three dimensions returns performances that both in terms of accuracy and computational requirements appear lower than those obtained with the other two grids.
Nevertheless, the \lognhg\ is the only grid among those considered in this paper that really allows to study the long-wavelength dynamics and, as discussed in the introduction, for such a study the \lognhg\ provides a clever choice also with respect to a purely logarithmic grid.

Interestingly enough, in a two-dimensional system the results show a similar behavior in terms of accuracy and the opposite behavior in terms of computational costs. 
For hard disks, integration on the \nhg\ leads to an accuracy of the critical packing fraction value which is similar to the one reached with standard integration but with the disadvantage that the former displays computational times which are about 5 times longer than the computational times required by the latter for the same size of the grid.
In contrast, the \lognhg\ displays an accuracy in determining the critical packing fraction which is comparable to the other two cases but with a lower computational effort for most of the grid sizes.
However, in the case of the \lognhg\, half of the grid points are placed in the logarithmic-grid part for $q < 3/\sigma$ and these might not be optimal.
Thus, to further optimize the efficiency of the numerical integration, one could in principle investigate other recipes in which the grid is constructed by taking into consideration a different fraction between the number of grid points placed in the logarithmic part and the number of grid points in the nonuniform part.

Both for $d=3$ and $d=2$, the evaluation of the nonergodicity parameters at different values of $M$ further demonstrates that the use of the \nhg\ should be preferred to the use of a uniform grid, and that also integration on the \lognhg\ returns good performance for the lowest values of $M$, in particular in the two-dimensional system.
Furthermore, in contrast to the case of standard integration on a uniform grid, integration on both the \nhg\ and the \lognhg\ returns nonergodicity parameters which show the correct behavior for $q\sigma \lesssim 1$, a range of wave numbers relevant to Brillouin and other light-scattering measurements.

The main conclusion of our study is that moderate but significant improvements in the performances can be obtained by using a non-uniform grid. 
Depending on the level of accuracy needed this may allow decreasing the grid size by a factor of two, which is particularly relevant in two dimensions where the computational effort required to solve MCT equations scales as the third power of the grid size.
This improvement may pave the way to a more systematic MCT investigation of more advanced topics as confined or non-homogeneous liquids, multi-component liquids, molecular liquids, and systems of active particles.
In fact, in all these cases matrix-valued correlation functions increase tremendously the computational costs and, even if one can circumvent the excessive computational times by parallelizing the MCT kernel evaluation, overcoming the memory allocation limits poses a more serious limitation.
Therefore, the reduction of wave numbers required for an accurate representation of the numerics is absolutely crucial in these cases.

Finally, we conclude by noting that also other integration schemes can be adopted independently of the grid type. 
For example, one can perform Simpson integration instead of Riemann or trapezoidal integration. 
Preliminary tests show that in the case of hard spheres, Simpson integration performs slightly better than the standard Riemann integration but with an obvious increase in the computational costs at equivalent grid size.
Finally, in the case of hard disks, the expression of the inner integral of the MCT kernel, see Eq.~(\ref{eq_appendixA1}), naturally suggests resorting to Chebyshev--Gauss integration.
However, preliminary results obtained on a uniform grid show that there is no accuracy improvement, likely due to the fact that Chebyshev nodes never coincide with the grid points and then one has always to approximate the various functions appearing in Eq.~(\ref{eq_appendixA1}), for example by using linear interpolation.

\section*{Acknowledgments}
We thank Matthias Fuchs for useful discussions. This work has been supported by the Austrian Science Fund (FWF): I 2887-N27 'Structure and Dynamics of Liquids in Confinement' 

%%Vancouver style references.
%\bibliographystyle{apsrev4-1}
\bibliographystyle{plain}
\bibliography{references}

\end{document}